\documentclass[pra,superscriptaddress,nofootinbib,twocolumn,showpacs]{revtex4-1}
\usepackage{amsmath,graphicx,epsfig,color,amsfonts,amssymb,bbm}
\usepackage{amsthm}
\usepackage{mathtools}

\newcommand{\id}{\mathbbm{1}}
\newcommand{\half}{\frac{1}{2}}
\newcommand{\vecP}{\vec{P}}

\newcommand{\ibid}{ {\em ibid.}}

\newcommand{\NS}{\mathcal{NS}}
\newcommand{\R}{\mathcal{R}}

\newcommand{\Q}{\mathcal  Q}
\newcommand{\T}{\mathcal  T}

\newcommand{\ket}[1]{|#1\rangle}
\newcommand{\bra}[1]{\langle#1|}
\newcommand{\braket}[1]{\left\langle #1 \right\rangle}
\newcommand{\proj}[1]{\left| #1 \right\rangle\!\!\left\langle #1 \right|}

\newcommand{\vecx}{{\vec{x}}}
\newcommand{\veca}{\vec{a}}

\newtheorem{theorem}{Theorem}
\newtheorem{corollary}{Corollary}

\usepackage{hyperref}

\renewcommand{\P}{\mathcal{P}}
\renewcommand{\S}{\mathcal{S}}
\renewcommand{\L}{\mathcal{L}}

\begin{document}

\title{Quantifying multipartite nonlocality via the size of the resource}

\author{Florian John Curchod}
\email{florian.curchod@icfo.es}
\affiliation{ICFO--Institut de Ci\`encies Fot\`oniques, 08860 Castelldefels (Barcelona), Spain.}
\author{Nicolas Gisin}
\affiliation{Group of Applied Physics, University of Geneva, CH-1211 Geneva 4, Switzerland.}
\author{Yeong-Cherng~Liang}
\email{yliang@phys.ethz.ch}
\affiliation{Institute for Theoretical Physics, ETH Zurich, 8093 Zurich, Switzerland.}

\date{\today}

\begin{abstract}
The generation of (Bell-)nonlocal correlations, i.e., correlations leading to the violation of a Bell-like inequality, requires the usage of a nonlocal resource, such as an entangled state. When given a correlation (a collection of conditional probability distributions) from an experiment or from a theory, it is desirable to determine the extent to which the participating parties would need to collaborate nonlocally for its (re)production. Here, we propose to achieve this via the {\em minimal group size} (MGS) of the resource, i.e., the smallest number of parties that need to share a given type of nonlocal resource for the above-mentioned purpose. In addition, we  provide a general recipe --- based on the lifting of Bell-like inequalities --- to construct MGS witnesses for non-signaling resources starting from {\em any} given ones. En route to illustrating the applicability of this recipe, we also show that when restricted to the space of full-correlation functions, non-signaling resources are as powerful as unconstrained signaling resources. Explicit examples of correlations where their MGS can be determined using this recipe and other numerical techniques are provided.
\end{abstract}

\maketitle

\section{Introduction} 
Quantum correlations that violate a Bell-type inequality~\cite{Bell1964}, a constraint that was first derived in the studies of local-hidden-variable-theories, were initially perceived only as a counterintuitive feature that has no classical analog. Following the discovery of quantum information science, these bizarre correlations have taken the new role as a resource. For instance, in the context of nonlocal games~\cite{Cleve2004} (which are closely related to the studies of interactive proof systems in complexity theory, see, eg. Ref.~\cite{Doherty:CCC:2008}), nonlocal, i.e., Bell-inequality-violating, correlations are those that cannot be simulated by shared randomness (SR). They are also well-known as an indispensable resource in quantum information processing and communication tasks such as the reduction of communication complexity~\cite{CommComplexity}, the distribution of secret keys in a device-independent setting~\cite{DIQKD}, as well as the certification and expansion of randomness~\cite{Randomness} etc. For a  comprehensive review on these and other applications, see Ref.~\cite{Brunner:RMP}.
 
As in any other resource theory~\cite{ResourceTheory}, the inter-convertibility of resources, and the possibility to substitute one by another in a certain task are important ingredients that put our understanding of these resources on a firm ground. Considerable effort has been devoted to these questions in the bipartite setting  --- in the cost of simulating quantum correlations using classical communication~\cite{TonerSimulation} or certain ``nonlocal boxes"~\cite{Cerf:PRL:2005}, as well as the inter-convertibility between these different resources~\cite{Barrett:PRA:2005,Jones:2005} (see Ref.~\cite{Brunner:RMP} for a review). However, relatively little  is known~\cite{Barrett:PRL:95,Multi:Simulation} in the multipartite scenarios, where other interesting features are also present, such as the monogamy of nonlocal correlations (see eg., Refs.~\cite{Brunner:RMP,Monogamy} and references therein)  and the possibility of them being anonymous~\cite{ANL}.

\begin{figure}[h!]
  \scalebox{0.52}{\includegraphics{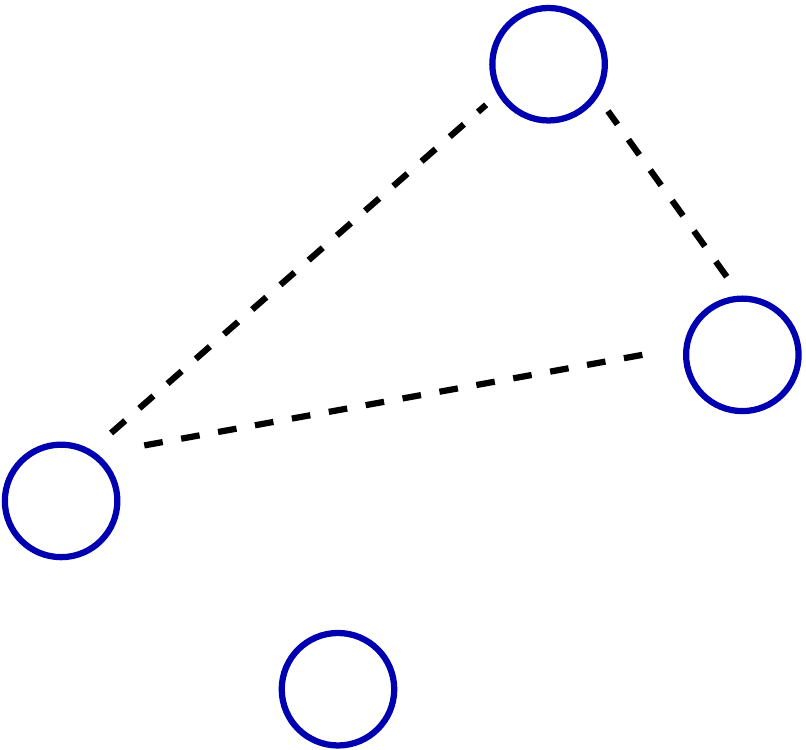}}
   \caption{\label{Fig:3-producible} Schematic diagram showing a situation where the conventional approach of non-biseparability fails to detect the multipartite nonlocality present in the correlation. Here, the dashed lines joining the three circles symbolically represent the nonlocal collaboration between the three parties. Since the fourth party is only correlated with the rest through shared randomness, the overall  correlation is biseparable. }
  \end{figure}

Thus far, prior investigations on multipartite nonlocal correlations have focused predominantly on their $m$-separability, namely, the possibility to reproduce them  when the parties are separated into $m$ groups~\cite{Bancal:PRL:2009} --- specifically two groups~\cite{Svetlichny,GMNL,Bancal:JPA:2012,Aolita:PRL:2012,Gallego:PRL:070401,Bancal:PRA:014102} --- and where the usage of some nonlocal resource $\R$ is only allowed within each group.\footnote{However, we do assume that global shared randomness is available for free in the resource theory of correlations.} While this has been a fruitful approach for the detection of genuine multipartite nonlocality, and hence genuine multipartite entanglement in a device-independent setting~\cite{DIEW,Pal:DIEW,Moroder:PRL:PPT,Bancal:JPA:2012}, it is however  not always applicable to the detection of genuine multipartite nonlocality among a {\em subset} of participating parties. To manifest this shortcoming, let us consider a 4-partite correlation $\vecP=\{P(\vec{a}|\vec{x})\}=\{P(a_1a_2a_3a_4|x_1x_2x_3x_4)\}$ of getting measurement outcome (output) $a_i$ for the $i$-th party given the measurement setting (input) $x_i$. A specific kind of biseparable correlation in this scenario takes the form of
\begin{equation}\label{Eq_bisep4_2vs2}
\begin{split}
	P(\vec{a}|\vec{x}) &= \sum_\lambda q_\lambda P^\R_\lambda(a_1a_2|x_1x_2)P^\R_\lambda(a_3a_4|x_3x_4)\\
	&+\sum_\mu q_\mu P^\R_\mu(a_1a_3|x_1x_3)P^\R_\mu(a_2a_4|x_2x_4)\\
	&+\sum_\nu q_\nu P^\R_\nu(a_1a_4|x_2x_3)P^\R_\nu(a_2a_3|x_2x_3),
\end{split}
\end{equation}
where $P^\R_i(a_ja_k|x_jx_k)$ is some  2-partite distribution allowed by the resource $\R$, while $q_\lambda$, $q_\mu$ and $q_\nu$ are non-negative, normalized weights. If  $\vecP$  {\em cannot} be written in the form of Eq.~\eqref{Eq_bisep4_2vs2},  the production of this correlation clearly requires at least 3 out of the 4 participating parties to collaborate nonlocally via $\R$.
If moreover,  nonlocal collaboration between 3 parties is sufficient, we see that $\vecP$ is thus biseparable, i.e., producible by parties separated into (convex mixtures of) two groups, see Fig.~\ref{Fig:3-producible}. In other words, the multipartite nonlocality contained in $\vecP$ cannot be detected by the conventional approach of detecting non-biseparability. Indeed, with the conventional ($m$-separability) approach, one only makes a distinction between the number of groups, but not the size, i.e., the number of parties involved in each group.

To determine the extent to which participating parties would need to collaborate nonlocally in a general scenario, it thus seems more natural to quantify multipartite nonlocality in terms of the {\em minimal group size} (MGS), i.e., the smallest number of parties required to collaborate nonlocally in reproducing some nonlocal correlation. Clearly, this approach provides information complementary to the one of $m$-separability on how $\R$  has to be distributed/ shared among the participating parties in order to reproduce some given correlation. The aim of this paper is to give a state-of-the-art exposition of this approach and to provide a general technique for the construction of MGS witnesses.

The rest of this paper is structured as follows. In Section~\ref{Sec:MGSetc}, we give a more formal introduction to the notion of MGS and the closely related concept of $k$-producibility; their connection to the conventional notion of genuine multipartite nonlocality is also discussed therein. Then in Section~\ref{Sec:Characterization}, we give an exposition of some basic facts about the sets of correlations that are $k$-producible. After that, in Section~\ref{Sec:ExampleFromNS22}--\ref{Sec:ExampleFromANL}, we give examples of quantum-realizable correlations where their characterization via the MGS approach is both natural and explicit. In Section~\ref{Sec:MGSFromBI}, we provide a general recipe to construct $n$-partite witness of non-$k$-producibility --- i.e., witnesses certifying MGS $> k$ --- starting from any given witness involving a smaller number of parties. There, we also make a digression to point out the universality of non-signaling resource when one is only concerned with the so-called full-correlation functions~\cite{Bancal:JPA:2012}. Finally, we conclude with some possible future research in Sec.~\ref{Sec:Conclusion}. Proofs of the two theorems and one corollary given in Section~\ref{Sec:MGSFromBI} are relegated to the Appendices.

\section{Minimal group size, $k$-producibility  and  multipartite nonlocality}
\label{Sec:MGSetc}

Formally, let us remind that an $n$-partite correlation $\vecP=\{P(\vec{a}|\vec{x})\}$ is a collection of the conditional probability distributions of getting  outputs  $\vec{a}=(a_1,a_2,\ldots, a_n)$, given the  inputs  $\vec{x}=(x_1,x_1,\ldots,x_n)$. In analogy with the studies of multipartite entanglement~\cite{Guehne:NJP:2005}, we say that $\vecP$ is $k$-producible (or more precisely $k$-partite $\R$-producible) if 
\begin{enumerate}
\item $\vecP$ can be decomposed into a convex mixture of products of {\em at most} $k$-partite correlations,   and
\item each  constituent correlation satisfies the constraints defined by the resource $\R$.\footnote{For example, if $\R$ refers to a quantum resource, then the constituent correlation must be producible by performing local measurements on some quantum state.}
\end{enumerate}
As a basic example, we note that by definition, a correlation $\vecP$ is Bell-local (henceforth local) if
\begin{equation}\label{Eq:locality}
	P(\vec{a} |\vec{x}) = \sum\limits_{\lambda}  q_{\lambda} \prod_{i=1}^n P_{\lambda}(a_i | x_i),
\end{equation}
for some choice of normalized weights $q_\lambda\ge0$ and some constituent correlations $P_\lambda(a_i|x_i)$. A local correlation is thus 1-producible,  and hence producible by  each party being alone and sharing no nonlocal resource with the others. On the other hand, correlation satisfying Eq.~\eqref{Eq_bisep4_2vs2} is 2-producible whereas a correlation $\vecP$ satisfying
\begin{equation}\label{Eq_bisep4_3vs1}
	\begin{split}
	P(\vec{a}|\vec{x}) &= \sum_\lambda q_\lambda P^\R_\lambda(a_1a_2a_3|x_1x_2x_3)P^\R_\lambda(a_4|x_4)\\
	&+\sum_\mu q_\mu P^\R_\mu(a_1a_2a_4|x_1x_2x_4)P^\R_\mu(a_3|x_3)\\
	&+\sum_\nu q_\nu P^\R_\nu(a_1a_3a_4|x_1x_3x_4)P^\R_\nu(a_2|x_2)\\
	&+\sum_\theta q_\theta P^\R_\theta(a_2a_3a_4|x_2x_3x_4)P^\R_\theta(a_1|x_1).
	\end{split}
\end{equation}	
is 3-producible. A general 3-producible correlation, however, may involve convex combination of correlation of the form of Eq.~\eqref{Eq_bisep4_2vs2} and of Eq.~\eqref{Eq_bisep4_3vs1}.
Obviously, $k$-producibility implies $k'$-producibility for all $k'>k$. Using the above terminologies, we thus say that $\vecP$ is {\em genuinely} $k$-partite nonlocal\footnote{\label{GMEvsGMNL}To conform with existing terminologies in the literature, when $\R$ refers to a quantum resource, we say that $\vecP$ must have arisen from a genuinely $k$-partite entangled state instead of $\vecP$ exhibits genuine $k$-partite nonlocality.} or having a MGS of $k$ if $\vecP$ is $k$-producible but not $(k-1)$- producible. For example, a 4-partite correlation that satisfies Eq.~\eqref{Eq_bisep4_2vs2} but not Eq.~\eqref{Eq:locality} is 2-producible but not 1-producible, and hence genuinely 2-partite nonlocal. Similarly, a 4-partite correlation that is 3-producible but not decomposable in the form of Eq.~\eqref{Eq_bisep4_2vs2} is genuinely 3-partite nonlocal.

A few other remarks are now in order. Firstly, the above definition can be seen as a generalization of existing notions of genuine $k$-partite nonlocality for an $n\!\!=\!\!k$-partite scenario~\cite{Bancal:PRA:014102} to an $n$-partite scenario where $n\ge k$. 
It is worth noting that the question of whether a given correlation $\vecP$ can be produced by having {\em at most} $k$ parties in one group ($k$-producibility) is not completely independent from the question of whether $\vecP$ can be produced by separating the $n$ parties into {\em at least} $m$ groups ($m$-separability).
For instance, a $k$-producible correlation $\vecP$ is $m$-separable  for some $m\ge  \lceil \frac{n}{k}\rceil$; likewise, if $\vecP$ is $m$-separable, it is also $k$-producible for some $k\ge \lceil \frac{n}{m}\rceil$. Thus, the smallest $n$ for which these descriptions become inequivalent is $n=4$. Finally, any multipartite correlation that cannot be produced by SR, or equivalently that is nonlocal [cf. Eq.~\eqref{Eq:locality}], or not 1-producible, is genuinely $k$-partite nonlocal for some $k\ge 2$.

\subsection{Characterization of the sets of $k$-partite $\R$-producible correlations}
\label{Sec:Characterization}

\begin{figure}[h]
  \scalebox{0.6}{\includegraphics{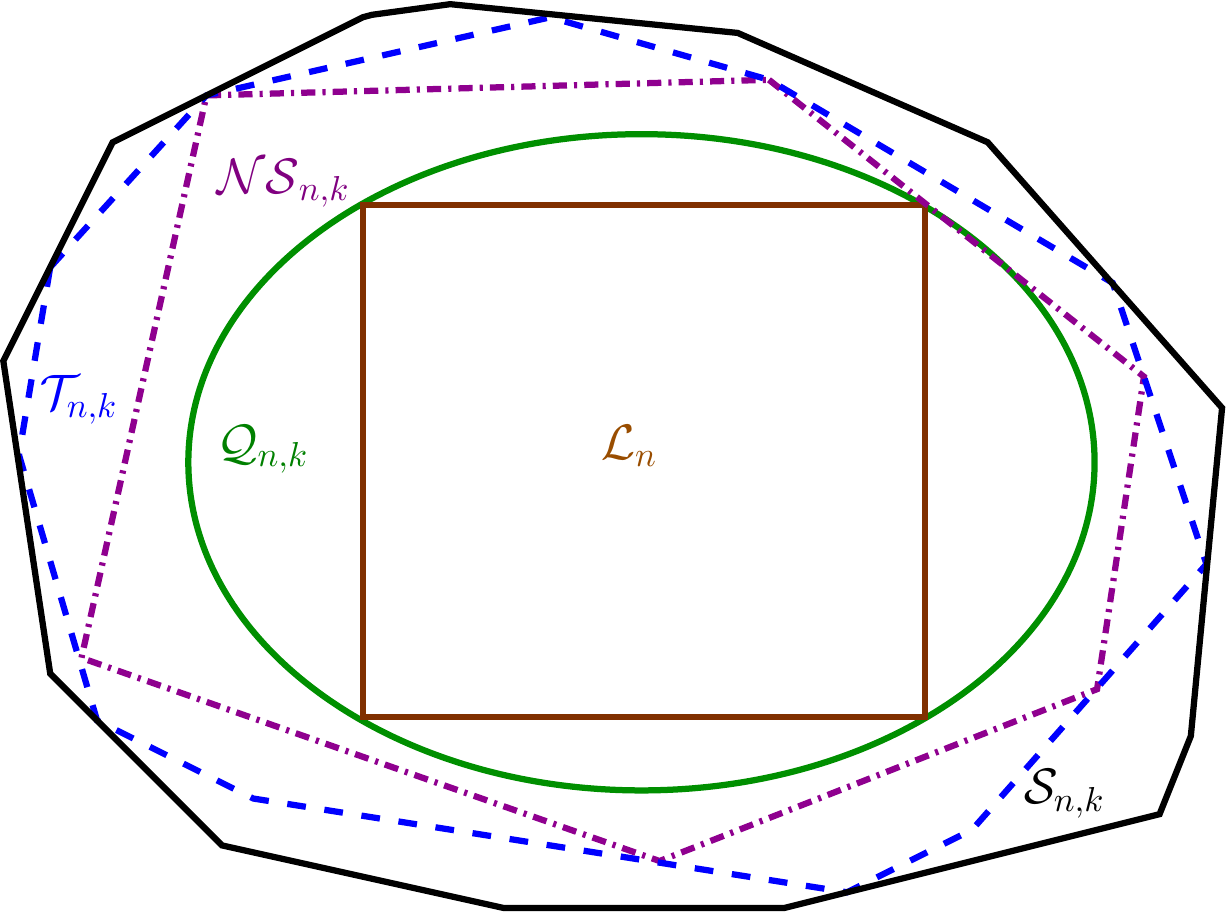}}
   \caption{\label{Fig:Hierarchy} (Color online) Schematic diagram showing the inclusion relations of the various sets of $\R_{n,k}$, cf. Eq.~\eqref{Eq:Inclusions} and Eq.~\eqref{Eq:Implications}. The smallest of these sets is $\mathcal{L}_n$ [depicted as the (brown) rectangle], followed by $\Q_{n,k}$ [depicted as the (green) oval], followed by $\NS_{n,k}$ [with boundary  marked by the (magenta)  dashed-dotted line], followed by $\T_{n,k}$ [with boundary marked by the (blue) dashed line]. Finally, the $k$-producible Svetlichny set $\S_{n,k}$ is represented by the outermost (black) solid polygon.   }
  \end{figure}

While the bulk of the above discussion is
 independent of the choice of  the nonlocal resource $\R$, it is worth reminding some features that are pertinent to specific resources. In this context, four commonly discussed nonlocal resources $\R$ are: (1) $\Q$: (local measurements on) an entangled quantum state of unrestricted Hilbert space dimension, (2) $\NS$: a post-quantum, but non-signaling~\cite{NS,Barrett:PRA:2005} resource,\footnote{Such a resource only allows correlations where their marginal distributions for {\em any} subset of parties are  independent of the input of the complementary subset of parties.} (3) $\T$~\cite{Gallego:PRL:070401,Bancal:PRA:014102}: a time-ordered, one-way classical signaling resource\footnote{The correlations allowed by such a resource is referred to as time-ordered bilocal in Ref.~\cite{Gallego:PRL:070401}.}  and (4) $\S$~\cite{Svetlichny} : a Svetlichny resource.\footnote{The Svetlichny resource allows the parties in a group to use any joint strategy and hence to produce any correlation that is only constrained by the normalization of probabilities. In some cases, such a resource can be realized by allowing multiple rounds of classical communications among the parties but in others, such a resource may not have a well-defined physical meaning, see Refs.~\cite{Gallego:PRL:070401,Bancal:PRA:014102} for a discussion.} Note that each resource $\R$ above is strictly stronger than the preceding one(s), in the sense that $\R$ can be used to produce all correlations arising from the preceding resource(s)~\cite{Gallego:PRL:070401,Bancal:PRA:014102}. As a result, we have the strict inclusion relations,
\begin{equation}\label{Eq:Inclusions}
	\mathcal{L}\subset\Q\subset\NS\subset\T\subset\S,
\end{equation}
with $\mathcal{L}$ being a local resource, provided by SR alone. Hence, a correlation $\vecP$ that is $k$-partite $\Q$-producible is also a member of the set of $k$-partite $\R$-producible correlations (henceforth denoted by $\R_{n,k}$) for $\R\in\{\NS,\T,\S\}$. Conversely, a correlation that is {\em not} in $\S_{n,k}$ is also not in $\R_{n,k}$ for $\R\in\{\Q,\NS,\T\}$, see Fig.~\ref{Fig:Hierarchy}. Formally, these implications are summarized as follows:
\begin{subequations}\label{Eq:Implications}
\begin{gather}
	\label{Eq:QimpliesOthers}
	\vecP\in\Q_{n,k}\Rightarrow \vecP\in \R_{n,k}\quad\text{for all}\quad\R\in\{\NS,\T,\S\},\\
	\vecP\not\in\S_{n,k}\Rightarrow \vecP\not\in \R_{n,k}\quad\text{for all}\quad\R\in\{\Q,\NS,\T\}.
	\label{Eq:SimpliesOthers}
\end{gather}
\end{subequations}

More generally, we note that independent of the nonlocal resource $\R\in\{\Q,\NS,\T,\S\}$, the set $\R_{n,k}$ is convex. Moreover, for the case when $\R\in\{\NS,\T,\S\}$,
$\R_{n,k}$ is even a convex polytope~\cite{Bancal:PRA:014102}, i.e., a convex set having only a finite number of extreme points~\cite{Polytope} and thus can be equivalently specified through a finite number of Bell-like inequalities (corresponding to the facets of the respective polytope). Determining if a given correlation $\vecP$ is inside $\R_{n,k}$, and hence producible by the respective resource can thus be decided via a linear program~\cite{ConvexOpt}, or through the violation of one of those Bell-like inequalities defining the polytope. In the simplest 2-input,  2-output scenario where $\R=\NS$, the set $\NS_{3,2}$ has been completely characterized in Ref.~\cite{Bancal:PRA:014102} whereas a superset of $\NS_{4,2}$ has also been characterized in Ref.~\cite{NS22} (see also Ref.~\cite{Curchod:Thesis}). 
If $\R=\Q$, i.e., a quantum resource, then the set $\R_{n,k}$ is no longer a convex polytope. Determining if a given $\vecP$ is in $\Q_{n,n-1}$ can nonetheless be achieved by solving a hierarchy of semidefinite programs~\cite{ConvexOpt} described in Ref.~\cite{DIEW}. More generally, determining if any given $\vecP$ is in $\Q_{n,k}$ can be achieved --- to some extent --- by solving a variant of the hierarchy of semidefinite programs described in Ref.~\cite{Moroder:PRL:PPT} (see Ref.~\cite{Liang:IP} for details).

However, regardless of $\R$, it is generally formidable to solve the aforementioned linear/ semidefinite programs by {\em brute force} even on a computer for relatively simple scenarios. Implications such as those summarized in Eq.~\eqref{Eq:Implications} are thus useful to bear in mind for subsequent discussions.  For example, if $\vecP$ violates an $n$-partite Svetlichny inequality --- a Bell-like inequality that holds for a general Svetlichny resource --- then it is not $(n-1)$-producible for all $\R$. In other words, the correlation $\vecP$  exhibits genuine $n$-partite nonlocality (and hence can only be produced, if at all, by a genuinely $n$-partite entangled state) and has an MGS of $n$. A generic correlation $\vecP$, evidently, will have an MGS  that depends on the resource under consideration, as we now illustrate by explicit examples in the following subsections.

\subsection{An example of a genuinely 3-partite $\NS$ nonlocal correlation in a 4-partite scenario}
\label{Sec:ExampleFromNS22}

The Greenberger-Horne-Zeilinger (GHZ) state~\cite{GHZ} between $n$ parties is defined as follow :
\begin{equation}\label{Eq:GHZ}
	\ket{\rm GHZ_n} = \frac{1}{\sqrt{2}}  (\ket{0}^{\otimes n}+\ket{1}^{\otimes n}),
\end{equation}
where $\ket{0}$ and $\ket{1}$ are, respectively, the eigenstate of the Pauli matrix $\sigma_z$ with eigenvalue $+1$ and $-1$. Consider the following  equal-weight mixture of three parties sharing $\ket{\rm GHZ_3}$ and one party holding $\ket{-}$:
\begin{equation}\label{Eq_rhoDN}
	\rho = \frac{1}{4} \left(\proj{\rm GHZ_3}\otimes\proj{-}+\circlearrowright\right) 
\end{equation}
where $\ket{-}$ is the eigenstate of the Pauli matrix $\sigma_x$ with eigenvalue $-1$, and we have used $\circlearrowright$ to denote similar terms which must be included to ensure that the expression involved is invariant under arbitrary permutation of parties. This quantum state could be prepared, for instance, by distributing uniformly randomly $\ket{\rm GHZ_3}$ to any of the three parties and  $\ket{-}$ to the remaining one. By construction, $\rho$ does not have genuine 4-partite entanglement. Hence, any correlation $\vecP$ derived by performing local measurement on $\rho$ must be a member of $\Q_{4,3}$ and by Eq.~\eqref{Eq:QimpliesOthers}, also $\R_{4,3}$.

Now, consider the case where all parties measure the following dichotomic observables,
\begin{equation}\label{Eq_ObsDN}
\begin{split}
A_0 = B_0=C_0=D_0=- \frac{\sqrt{3}}{2}\sigma_x+\half\sigma_y,\\
A_1 = B_1=C_1=D_1= - \frac{\sqrt{3}}{2}\sigma_x-\half\sigma_y.
\end{split}
\end{equation}
It can be shown that that the resulting correlation $\vecP$ violates the following Bell inequality which must be satisfied by all correlations from $\NS_{4,2}$~\cite{NS22}:
\begin{align}
\mathcal{I}= &-12  \braket{A_0}	-3   \braket{A_1}	-2  \braket{A_0 B_0} +6  \braket{A_0 B_1} \nonumber\\
&-3  \braket{A_1 B_1}+13  \braket{A_0 B_0 C_0}   -3  \braket{A_1 B_0 C_0}	\nonumber\\
&-11  \braket{A_1 B_1 C_0}+14  \braket{A_1 B_1 C_1}+22  \braket{A_0 B_0 C_0 D_0}	\nonumber\\
&-15  \braket{A_0 B_0 C_0 D_1}-10  \braket{A_1 B_1 C_0 D_0}\nonumber\\
&-7  \braket{A_1 B_1 C_1 D_0}+21  \braket{A_1 B_1 C_1 D_1}+	\circlearrowright \nonumber\\
&\stackrel{\NS_{4,2}}{\le} 105,\label{Ineq_DN}
\end{align}
giving a quantum value of 117.8827. This implies that the correlation $\vecP$ is also genuinely 3-partite nonlocal, or having an MGS of 3 for $\R\in\{\Q,\NS\}$. 

Interestingly, it can be shown that $\vecP$ {\em does not} lie in any of the 3-partite $\NS$-producible set corresponding to a fixed partition. This, together with the fact that $\vecP$ is 3-partite $\NS$-producible means that the generation of $\vecP$ requires classical mixtures of different partitions of the 4 participating parties into 2 groups, one of them containing three parties and sharing an $\NS$ resource.
It is also worth noting that all  {\em tripartite} marginal correlations of $\vecP$ are verifiably 1-producible (hence satisfying the complete set of Bell inequalities for this scenario given in Ref.~\cite{Sliwa}). In other words, although $\vecP$ is genuinely 3-partite $\NS$-nonlocal, this 3-partite nonlocality cannot be revealed by studying each of the four tripartite marginal correlations individually. Neither can this multipartite nonlocality be manifested by analyzing the biseparability of the 4-partite correlation since this more conventional approach cannot distinguish correlation of the form of Eq.~\eqref{Eq_bisep4_2vs2} and those of the form of Eq.~\eqref{Eq_bisep4_3vs1}.

In the above example, we were able to determine the MGS of the correlation for  the quantum, and a general non-signaling resource. For the Svetlichny resource, we could also show --- by solving some  linear program --- that the very same correlations is inside the set $\S_{4,2}$, and thus only exhibits an MGS of 2. However, due to the computational complexity involved in solving the corresponding linear program for the 1-way signaling resource $\T$, we were not able to determine precisely its MGS.  Apart from correlations that violate an $n$-partite Svetlichny inequality (in which case MGS  $=n$ for all resources)  or correlations that are local  (in which case MGS $=$ 1), one may thus wonder if there exist other $n$-partite correlations $\vecP$  which have an MGS that can be fully characterized for all the different resources. We now provide examples of this kind in the next section.

\subsection{A family of $n$-partite examples with fully characterized MGS} 
\label{Sec:ExampleFromANL}

In Ref.~\cite{ANL}, it has been shown that if all $n$ parties either measure the $\sigma_x$ or the $\sigma_y$ observable on the $n$-partite state $\ket{\rm GHZ_n}$, Eq.~\eqref{Eq:GHZ}, the resulting correlation has an MGS of $\lceil \frac{n}{2}\rceil$ for $\R\in\{\NS,\T,\S\}$ whenever $n$ is odd or $\frac{n}{2}$ is even. On the other hand, if we restrict ourselves to a quantum resource, then for all odd $n\ge 3$, it follows from the result of Ref.~\cite{ANL} that the corresponding MGS is $n$, demonstrating a large gap between the size of the resource required to reproduce these correlations when using a quantum and a post-quantum non-signaling (or a classical but signaling) resource. To prove these results, a general $\NS$ biseparable decomposition of the aforementioned correlation was provided~\cite{ANL} for arbitrary partitioning of the $n$ parties into two groups, thus establishing that these correlations are $\lceil \frac{n}{2}\rceil$-producible for $\R\in\{\NS,\T,\S\}$. Then to prove that these correlations are {\em not} $(\lceil \frac{n}{2}\rceil-1)$-producible for the same set of resources, it was shown in Ref.~\cite{ANL} that except for even $n$ with odd $\frac{n}{2}$, these correlations are not 3-separable, i.e., cannot be reproduced by a separation of the $n$ parties into 3 groups. As for $\R=\Q$, an MGS = $n$ for odd $n$~\cite{ANL} follows from the fact that the corresponding correlation violates a device-independent witness for genuine $n$-partite entanglement~\cite{DIEW,Bancal:JPA:2012} constructed from the Mermin-Ardehali-Belinskii-Klyshko (MABK) Bell expression~\cite{MABK1,MABK2}. In the case of even $n$,  result recently established in Ref.~\cite{Liang:IP} (based on earlier work of Ref.~\cite{Nagata:PRL}) allows one to conclude that the above-mentioned GHZ correlations has an MGS of at least $n-1$ (for $\R=\Q$).

\section{Witnessing non-$k$-producibility using Bell-like inequalities}
\label{Sec:MGSFromBI}

Evidently, as discussed in Sec.~\ref{Sec:MGSetc}, Bell-like inequalities are very useful tools for determining (or at least lower-bounding) the MGS of a given correlation by certifying its non-$k$-producibility. For example, all Bell-like inequalities that have been derived --- based on the non-biseparability approach~\cite{Svetlichny,GMNL,Bancal:JPA:2012,Aolita:PRL:2012, Gallego:PRL:070401,Bancal:PRA:014102} --- to detect genuine $n$-partite nonlocality can be used as witnesses for non-$(n-1)$-producibility for the respective resources. It is however unrealistic to hope to find {\em all} such Bell-like inequalities by solving the polytope describing the convex set $\R_{n,k}$ even for relatively small $n$ and $k$. But all is not lost and in this section, we recall from Ref.~\cite{Pironio:Lifting} the technique of lifting --- originally developed for Bell inequalities that witness Bell-nonlocality --- and show that it can also be used to construct Bell-like inequality for arbitrary $\R_{n',k}$ (where $n'>n$ and $\R\in\{\Q,\NS\}$) starting from {\em any} given Bell-like inequality for $\R_{n,k}$. Before that, let us first make a digression and point out the usefulness of a non-signaling resource in simulating a general correlation.

\subsection{All extremal full-correlation functions can be simulated with non-signaling strategies}

In an $n$-partite, $m$-input, $\ell$-output Bell scenario, the set of full-correlation functions defined in Ref.~\cite{Bancal:JPA:2012} consists of the following $\ell\, m^n$ joint conditional probability distributions:
\begin{subequations}\label{Eq:FullCorrelatorSpace}
\begin{equation}\label{Eq:FullCorrelators}
	\{P([{\bf a}_\vecx]_\ell = r)\}_{r=0}^{\ell-1}
\end{equation}
where $[X]_\ell:= X\mod\,\ell$, 
\begin{equation}\label{Eq:Dfn:FullCorrelators}
	P([{\bf a}_\vecx]_\ell = r)=\sum_{\veca} P(\veca|\vecx)\,\delta_{\sum_i a_i\,{\rm mod}\, \ell,\, r},
\end{equation}
\end{subequations}
and $\delta_{a,b}$ is the Kronecker delta of $a$ and $b$. Note that due to the normalization conditions, only $(\ell-1)\, m^n$ of these joint conditional probability distributions are independent. Moreover, in the case where there are only two possible outcomes, i.e., $\ell=2$, it is easy to see that the above definition of full-correlation functions is equivalent to the conventional one defined by the expectation value of the product of $\pm1$ outcomes. 

We now present a mathematical fact about the space of correlations spanned by the set of full-correlation functions defined in Eq.~\eqref{Eq:FullCorrelatorSpace}.
\begin{theorem}\label{Thm:Simulation}
When restricted to the set of full-correlation functions given in Eq.~\eqref{Eq:FullCorrelatorSpace}, all extremal strategies achievable by an $n$-partite Svetlichny resource $\S_n$ are also achievable using an $n$-partite non-signaling resource $\NS_n$. Thus, in the subspace spanned by full-correlation functions, the three sets of correlations $\S_n$, $\T_n$, and $\NS_n$ become identical.
\end{theorem}

One can find the proof of this theorem in Appendix~\ref{App:FullCorrelators}. Let us remind that the Svetlichny resource is the most powerful nonlocal resource, and is only constrained by the normalization of probability distributions. In other words, $\S_n$ is basically the set of normalized $n$-partite correlations. The importance of Theorem~\ref{Thm:Simulation} is that when restricted to the set of coarse-grained measurement statistics represented by the set of full-correlation functions, cf Eq.~\eqref{Eq:FullCorrelators}, one also cannot make a distinction between $\NS_n$ and the set of normalized conditional probability distributions. Note that for binary-outcome full-correlation functions arising from the Bell singlet state, the coincidence between $\S_2$ and $\NS_2$ was already anticipated from the results of Ref.~\cite{Cerf:PRL:2005}. In fact, an alternative proof of Theorem~\ref{Thm:Simulation} for the special case of binary-outcome full-correlation functions can be found, e.g., in Theorem 12 of Ref.~\cite{Marco:Thesis}.

It is also worth noting that the definition of full-correlation functions is not unique, nevertheless, numerous Bell-like inequalities can be written in terms of the correlation functions defined in Eq.~\eqref{Eq:FullCorrelatorSpace}, see e.g., Ref.~\cite{Bancal:JPA:2012}. In this regard, note also the following corollary of Theorem~\ref{Thm:Simulation}, which allows us to relate Bell-like inequalities for $\NS_{n,k}$ with those of $\R_{n,k}$ for $\R\in\{\T,\S\}$.
\begin{corollary}\label{Cor:SameBound}
Let $\mathcal{I}^\R_{n,k}$ be a tight, full-correlation Bell-like inequality that holds for $\R\in\{\T,\S\}$, i.e.,
\begin{equation}\label{Ineq:Rn-k}
	\mathcal{I}^\R_{n,k}: \sum_{\vec{x}}\sum_{r=0}^{\ell-1} \beta^{r}_{\vecx} P([{\bf a}_\vecx]_\ell = r)\stackrel{\R_{n,k}}{\le} B^\R_{n,k},
\end{equation}
and there exists $P(\veca|\vecx)\in\R_{n,k}$ such that inequality~\eqref{Ineq:Rn-k} becomes an equality, then there also exists $P(\veca|\vecx)\in\NS_{n,k}$ such that inequality~\eqref{Ineq:Rn-k} becomes an equality. In other words, 
\begin{equation}\label{Ineq:Rn-k:NS}
	\sum_{\vec{x}}\sum_{r=0}^{\ell-1} \beta^{r}_{\vecx} P([{\bf a}_\vecx]_\ell = r)\stackrel{\NS_{n,k}}{\le} B^\R_{n,k},
\end{equation}
is also a {\em tight}, full-correlation Bell-like inequality that holds for $\R=\NS$.
\end{corollary}
The proof of the above corollary can be found in  Appendix~\ref{App:ProofCorollary}. The corollary tells us that if we restrict ourselves to Bell-like inequalities that only involve linear combination of  full-correlation functions, Eq.~\eqref{Eq:FullCorrelatorSpace},  then we cannot distinguish between correlations that are $k$-producible with respect to any of the resource $\R\in\{\NS,\T,\S\}$. In other words, for any given $n$ and $k$ and in the subspace of measurement statistics spanned by the set of full-correlation functions, cf. Eq.~\eqref{Eq:FullCorrelatorSpace}, the three sets of correlations $\NS_{n,k}$, $\T_{n,k}$ and $\S_{n,k}$ become identical. It is worth bearing this fact in mind in order to appreciate  the generality of the upcoming theorem.

\subsection{Lifting of Bell-like inequalities}

The {\em lifting} of Bell inequalities was first discussed by Pironio in Ref.~\cite{Pironio:Lifting}. Essentially, it is a technique that allows one to extend any (facet-defining) Bell inequality of a given scenario to a more complex scenario (involving more parties and/or inputs and/or outputs).
In this work, we are only interested in the lifting of  Bell-like inequalities to a scenario involving more parties. In this case, a lifted Bell inequality corresponds to a witness of nonlocality where the nonlocal behavior of a subset of, say $n$, of the parties becomes apparent after conditioning 
on a specific combination of measurement settings and outcomes from the complementary subset of $h$  parties.\footnote{This particular kind of lifting has been applied to show, for instance, a stronger version of Bell's theorem, see, e.g., Ref.~\cite{HiddenInfluences}.}

More concretely, let us denote a specific combination of the measurement settings and measurement outcomes of the $h$ parties, respectively, by $\vec{s}$ and $\vec{o}$. It can then be shown that if the $(n+h)$-partite correlation $P(\vec{a},\vec{o}|\vec{x},\vec{s})$ is 1-producible (and non-vanishing\footnote{If the distribution vanishes, the conditional distribution given in Eq.~\eqref{Eq:Conditional} is ill-defined.}), so is the conditional distribution given by:
\begin{equation}\label{Eq:Conditional}
	\tilde{P}^{|\vec{o},\vec{s}}(\vec{a} | \vec{x}) =  \frac{P(\vec{a},\vec{o}|\vec{x},\vec{s})}{\sum_{\vec{a}} P(\vec{a},\vec{o}|\vec{x},\vec{s})}.
\end{equation}
An immediate implication of this is that a Bell inequality that is defined for an $n$-partite scenario can be trivially extended to any $(n+h)$-partite scenarios by considering specific measurement settings $\vec{s}$ and outcomes $\vec{o}$ for the $h$ parties.

As an example consider the well known Clauser-Horne-Shimony-Holt~\cite{CHSH} Bell inequality applicable to a scenario involving two parties, each performing two binary-outcome measurements:
\begin{equation} \label{Eq:CHSH}
	\sum_{x_1,x_2,a_1,a_2=0}^1\!\!\! (-1)^{a_1+a_2 +x_1x_2}P(a_1a_2|x_1x_2)\stackrel{\L}{\le} 2.
\end{equation}
Lifting  this inequality to the scenario of 3 parties and with the 3rd party getting a {\em specific} measurement outcome $o_3$ given the {\em specific} measurement setting $s_3$ gives the following lifted CHSH Bell inequality:
\begin{equation} \label{Eq:LiftedCHSH}
\begin{split}
	\sum_{x_1,x_2,a_1,a_2=0}^1\!\!\!\!\!\! &(-1)^{a_1+a_2 +x_1x_2}P(a_1a_2o_3|x_1x_2s_3)\\
	&-2P(o_3|s_3)\stackrel{\L}{\le} 0. 
\end{split}
\end{equation}
Lifting the CHSH Bell inequality to an arbitrary number of $n>2$ parties can be carried out analogously. In Ref.~\cite{Pironio:Lifting}, it was shown that such a procedure not only generates a legitimate Bell inequality but even one that preserves the facet-defining property of the original Bell inequality.

\subsection{A general recipe for the construction of non-$k$-producible witnesses}

We shall now demonstrate how lifting may be used as a general technique for the construction of Bell-like inequalities for $\R_{n',k}$ starting from one for $\R_{n,k}$ where $n'$ is an arbitrary integer greater than $n$ and $\R$ is a resource that respects the non-signaling constraints.
To this end, we note that, without loss of generality, a (linear) Bell-like inequality for a non-signaling-respecting $\R_{n,k}$  can always be written in the form of:
\begin{equation}\label{Eq:GeneralBI}
	I_n = \sum_{\vec{a},\vec{x}} \beta^{\veca}_{\vecx} P(\vec{a}|\vec{x})\stackrel{\R_{n,k}}{\le} 0,
\end{equation}
where $\beta^{\veca}_{\vecx}$ is some real-valued function of $\veca$ and $\vecx$.
Our main observation is that the lifting of $I_n$ to a scenario involving arbitrary $n'>n$ parties is also a legitimate Bell-like inequality for $\R_{n',k}$, as summarized more formally in the following theorem.

\begin{theorem}\label{Thm}
If $I_{n}$ is a Bell-like inequality satisfied by all correlations in $\R_{n,k} \in \{\Q,\NS\}$, i.e., Eq.~\eqref{Eq:GeneralBI} holds for all $P(\veca|\vecx)\in\R_{n,k}$, then
\begin{equation}
	I_{n+h} = \sum_{\veca,\vecx} \beta^{\veca}_{\vecx} P(\veca,\vec{o}|\vecx,\vec{s})\stackrel{\R_{n+h,k}}{\le} 0,
\end{equation}
meaning that the lifted inequality holds for all $P(\veca,\vec{o}|\vecx,\vec{s})\in\R_{n+h,k}$ where $h\ge1$, whilst $\vec{o}$ and $\vec{s}$ refer, respectively, to arbitrary but {\em fixed} combination of measurement outcomes and measurement settings for the $h$ additional parties.
\end{theorem} 
A proof of this theorem can be found in Appendix~\ref{App:Proof}. Clearly, one can see Theorem~\ref{Thm} as a partial generalization of the results presented in~\cite{Pironio:Lifting} from $\R_{n,1}$ to $\R_{n,k}$ whenever $\R\in\{\Q,\NS\}$. As for $\R\in\{\T,\S\}$, we know from Corollary~\ref{Cor:SameBound} and Theorem~\ref{Thm} that any full-correlation Bell-like inequality valid for $\R_{n,k}$ can also be lifted as a Bell-like inequality for $\NS_{n',k}$ in the extended scenarios.
 Unfortunately, the theorem in general does not apply to the signaling resource  $\S$ (as well as $\T$). To see this, consider the tripartite Svetlichny inequality (writtten in the form given in~\cite{Bancal:JPA:2012}):
\begin{subequations}\label{Ineq_Svet3}
\begin{gather}
	I_{\S,3}=\sum_{\vecx,\veca} \beta^{\veca}_{\S,3,\vecx}\,P(a_1a_2a_3|x_1x_2x_3)-4\stackrel{\S_{3,2}}{\le} 0,\\
	\beta^{\veca}_{\S,3,\vecx}=(-1)^{\sum_i a_i+\left\lfloor \frac{\sum_i x_i-1}{2} \right\rfloor}.
\end{gather}
\end{subequations}
If Theorem~\ref{Thm} were to be applicable for a Svetlichny resource, we would expect, for instance, that the following inequality 
\begin{align}
	I_{\S,4}=&\sum_{\vecx,\veca} \beta^{\veca}_{\S,3,\vecx}\,P(a_1a_2a_3,o_4=0|x_1x_2x_3,s_4=0)\nonumber\\
	-&4\sum_{\veca}P(a_1a_2a_3,o_4=0|x_1'x_2'x_3',s_4=0)\le 0,
	\label{Ineq_Svet4}
\end{align}
to hold true for $\S_{4,2}$ and for some arbitrary choice of $x_1',x_2',x_3'=\{0,1\}$. One can, however, easily verify that this is not the case. For instance, with $x_1'=x_2'=x_3'=0$, the Svetlichny strategy from $\S_{4,2}$:
\begin{equation}
\begin{split}
	a_1&=1-\delta_{x_1,1}\delta_{x_2,1},\quad  a_2=1,\\
	a_3&=a_4=1-\delta_{x_3,1}\delta_{x_4,0},
\end{split}
\end{equation}
gives vanishing contribution to the second term in Eq.~\eqref{Ineq_Svet4} but an overall value of 4 for $I_{\S,4}$, clearly violating inequality~\eqref{Ineq_Svet4}.

Despite the above remark, let us stress once more that there is still wide applicability of Theorem~\ref{Thm}.
For example, {\em each} of the Bell-like inequalities obtained for $\NS_{3,2}$ and $\NS_{4,2}$ in Refs.~\cite{Bancal:PRA:014102,NS22} can now be used to construct witnesses showing MGS $\ge3$ (for the $\NS$ resource) for arbitrary number of parties. Thanks to Corollary~\ref{Cor:SameBound}, the families of  $k$-partite Svetlichny inequalities obtained in Refs.~\cite{GMNL,Bancal:JPA:2012} can similarly be extended to detect genuine $\NS$ $k$-partite nonlocality in an arbitrary $n>k$ partite scenario.
Likewise, {\em each} device-independent witness for genuine $k$-partite entanglement obtained in Ref.~\cite{DIEW,Bancal:JPA:2012,Pal:DIEW} can now be applied to witness genuine $k$-partite entanglement in an arbitrary $n>k$ partite scenario. Of course, it remains to show that Bell-like inequalities generated with the help of Theorem~\ref{Thm} could indeed be useful, and this is what shall show next with a very simple example.

\subsection{An example where a lifted Bell-like inequality can be used to determine MGS} 
\label{Sec:LiftedExample}

Consider the following four-partite mixed state:
\begin{align}\label{liftstate}
	\rho = v \proj{\rm GHZ_3} \otimes \proj{0} + (1-v) \frac{\id}{2^3} \otimes \ket{1}\bra{1},
\end{align}
where $v\in(0,1]$, and $\ket{0}$, $\ket{1}$ are again the eigenstates of $\sigma_z$. Since $\rho$ is {\em biseparable}, regardless of which local measurements are performed on $\rho$, the resulting correlations must be in $\Q_{4,3}$ and thus having MGS $\le 3$  for all $\R$ [cf. Eq.~\eqref{Eq:QimpliesOthers}]. Clearly, from Eq.~\eqref{liftstate}, we see that the entanglement of $\rho$ lies entirely within the first three subsystems. Let us denote these systems by $A$, $B$, and $C$ respectively. For $v\le \frac{1}{5}$, it is known that the tripartite reduced density matrix  $\rho_{\mbox{\tiny ABC}}=v \proj{\rm GHZ_3} + (1-v) \frac{\id}{2^3}$ is separable~\cite{GHZ:Sep} and thus not capable of violating any Bell inequalities. Nonetheless, in what follows, we shall show that a lifted Bell-like inequality can indeed be used to show that certain correlation derived from $\rho$ indeed exhibits MGS $= 3$ for all $v\neq0$, thus showing that the generation of such a correlation quantum mechanically indeed requires at least tripartite entanglement.

To this end, consider now the following dichotomic observables,
\begin{gather}
A_0 = \sigma_x,\quad A_1 = \sigma_y ,\nonumber \\ 
B_0 = \frac{1}{\sqrt{2}} (\sigma_x - \sigma_y) ,\quad B_1 =\frac{1}{\sqrt{2}} (\sigma_x + \sigma_y),
\label{Eq:SvetMeasurements}\\ 
C_0 = -\sigma_y , \quad C_1 = \sigma_x,\nonumber 
\end{gather}
and the tripartite Svetlichny inequality given in Eq.~\eqref{Ineq_Svet3}.
It is known that by measuring the local observables $\{A_i, B_i, C_i\}_{i=0,1}$ given in  Eq.~\eqref{Eq:SvetMeasurements}  on $\ket{\rm GHZ_3}$, one obtains correlation $\vecP$ that violates $I_{\S,3}$ maximally.

Note that by Corollary~\ref{Cor:SameBound} and the fact that Eq.~\eqref{Ineq_Svet3} is a full-correlation Bell-like inequality, we know that inequality~\eqref{Ineq_Svet3} still holds and can be saturated even if we now consider only correlations in $\NS_{3,2}$, i.e.,  
\begin{gather}\label{Ineq_NS3}
	I_{\NS,3}=\sum_{\vecx,\veca} \beta^{\veca}_{\S,3,\vecx}\,P(a_1a_2a_3|x_1x_2x_3)-4\stackrel{\NS_{3,2}}{\le} 0.
\end{gather}
Lifting the inequality $I_{\NS,3}$ to the specific case where the 4th party performs the 0-th measurement and getting the 0-th outcome, one obtains the inequality:
\begin{align}
	I^{|s_4=o_4=0}_{\NS,3}=&\sum_{\vecx,\veca} \beta^{\veca}_{\NS,3,\vecx}P(a_1a_2a_3,o_4=0|x_1x_2x_3,s_4=0)\nonumber\\
	&-4P(o_4=0|s_4=0)\stackrel{\NS_{4,2}}{\le} 0
	\label{Ineq_LiftedSvet}.
\end{align}
Let us now identify the 0-th measurement of the fourth party by $\sigma_z$ and the 0-th outcome by a successful projection onto the eigenstate $\ket{0}$. Together with the measurements specified in Eq.~\eqref{Eq:SvetMeasurements}, one finds that for all $0<v\le1$, the resulting correlation derived from $\rho$ must also violate inequality~\eqref{Ineq_LiftedSvet}. To see this, it suffices to note that (i) for $v\neq0$, the probability of successfully projecting the fourth system onto $\ket{0}$  is strictly greater than zero and (ii) conditioning on a successful projection, the conditional state for $ABC$  is simply $\ket{\rm GHZ_3}$ which, as mentioned above, violates $I_{\NS,3}$ inequality maximally. The aforementioned correlation thus exhibits MGS stronger than that allowed in $\NS_{4.2}$ which, by Eq.~\eqref{Eq:SimpliesOthers}, implies that it has MGS $\ge3$ for all $\R\in\{\Q,\NS\}$. Combining this with the biseparability of $\rho$ mentioned above, we see that this particular correlation has exactly MGS = 3.

\section{ Conclusion}
\label{Sec:Conclusion}

To investigate the extent to which participating parties would need to collaborate nonlocally in a nonlocal game (or equivalently in a Bell-type experiment), we have introduced the notion of {\em minimal group size} (MGS), i.e., the smallest number of nonlocally-correlated parties required to reproduce a given nonlocal correlation $\vecP$. We believe that this more general notion of genuine multipartite nonlocality inspired by $k$-producibility~\cite{Guehne:NJP:2005} from the studies of multipartite entanglement will be a fruitful approach towards a better understanding of multipartite nonlocality. 

As an illustration, we presented, in a four-partite scenario, some genuine tripartite nonlocal correlation where the multipartite nonlocality {\em cannot} be detected through the conventional $m$-separability approach. Nonetheless, as first demonstrated in Ref.~\cite{ANL}, and further elaborated in this paper, the biseparability approach can in some cases provide tight lower bound on MGS. In fact, for the family of $n$-partite correlations presented in Ref.~\cite{ANL}, it was even found that their MGS for a quantum resource is $n$ whereas that for a general non-signaling (or even an unrestricted signaling) resource is $\lceil \frac{n}{2}\rceil$, giving an increasing gap between their MGS as $n$ increases. Could there be a bigger gap between the MGS of a nonlocal correlation with respect to a quantum resource and a general (non-)signaling resource? In particular, does there exist a multipartite nonlocal quantum correlation which requires genuine $n$-partite entangled state for its production but nonetheless only an MGS of 2 if one is allowed to exploit a signaling, or even a non-signaling but post-quantum resource?  The answer to these questions would certainly shed light on how quantum entanglement help in a different aspect of communication complexity, namely, how many communicating parties we can replace by quantum entanglement.

We also demonstrated how the technique of lifting~\cite{Pironio:Lifting} --- originally presented in the context of Bell inequality (for 1-producibility) --- can be applied to generate new MGS witnesses starting from one involving a smaller number of parties. This generalizes partially the result of Ref.~\cite{Pironio:Lifting} and provides a useful recipe for the construction of MGS witnesses (with respect to a non-signaling, e.g., a quantum resource) for an arbitrary $n$-partite scenario. Moreover, we have found that for the complete list of 185 facet-defining Bell-like inequalities of $\NS_{3,2}$ given in Ref.~\cite{Bancal:PRA:014102}, the corresponding MGS witnesses of $\NS_{4,2}$ generated from lifting still correspond to a {\em facet}~\cite{Polytope} of the polytope in the more complex scenario. Likewise, when these 185 lifted inequalities, as well as the 13,479 facet-defining inequalities obtained in Ref.~\cite{NS22} are lifted  to the 5-partite scenario, it can again be verified that they correspond to facets of the $\NS_{5,2}$ polytope. Based on these observations, we conjecture that --- as with standard Bell inequalities --- the procedure of lifting, when applied to a facet-defining inequality of $\NS_{n,k}$, also generates a facet of $\NS_{n',k}$ in the extended scenario involving $n'>n$ parties.

Unfortunately, a naive application of lifting to signaling resources generally does not always result in legitimate MGS witnesses in the extended scenario. Nevertheless, the possibility to simulate all possible {\em full-correlation functions}~\cite{Bancal:JPA:2012} using only non-signaling resources --- as we show in Appendix~\ref{App:FullCorrelators} --- allows us to apply the recipe to Bell-like inequalities originally derived for Svetlichny resources~\cite{Bancal:JPA:2012,Svetlichny,GMNL} and construct MGS witnesses for non-signaling resources in {\em any} extended scenario. It is also  conceivable that an analogous witness-generating technique may be found for signaling resources, a problem that we shall leave for future research.

Evidently, on top of Bell-like inequalities that one may construct using the aforementioned technique, it is natural to ask if there exist simple family of non-$k$-producible witnesses for arbitrary number of parties. In this regard, we note that a family of such witnesses for a quantum resource (as well as a general non-signaling resource) has recently been identified~\cite{Liang:IP}. Similar results for other resources, especially one that is either {\em optimal} (in the sense of being facet-defining) for the respective convex polytope, or one that involves a small number of terms to be measured experimentally, would certainly be desirable.

Finally, let us stress that while we have discussed MGS mostly in the context of reproducing certain nonlocal correlations, these values for the post-quantum non-signaling resource, as well as for signaling resources also provide insight on the difficulty in reproducing certain correlations using quantum resources. In this sense, evaluation of the MGS for a given correlation may give an indication on how difficult it is to produce certain Bell-inequality violating correlations in the laboratory: the larger the value of MGS, the more systems need to be entangled together in their generation.

\begin{acknowledgments}
We acknowledge many stimulating discussions with Stefano Pironio, especially on suggesting a primitive form of the example given in Sec.~\ref{Sec:LiftedExample}. We are also grateful to Jean-Daniel Bancal for discussions and for his comments on an earlier version of this manuscript. This work is supported by the Swiss NCCR ``Quantum Science and Technology", the CHIST-ERA DIQIP and the ERC grant 258932. FJC acknowledges support from the John Templeton Foundation. 
\end{acknowledgments}

\appendix

\section{Proof that all full-correlation functions are attainable using a non-signaling resource}
\label{App:FullCorrelators}

Our goal here is to give a proof that when restricted to the $(\ell-1)\, m^n$-dimensional space of full-correlation functions defined by Eq.~\eqref{Eq:FullCorrelatorSpace}, the set of legitimate correlations coincide with that achievable by a non-signaling resource $\NS_n$. To this end, it is worth reminding that the set of normalized correlations in this space is precisely the set of correlations achievable by the Svetlichny resource $\S_n$. To prove the desired result, it is then sufficient to show that all extreme points of $\S_n$ in this space are also achievable using $\NS_n$.

\begin{proof}
Firstly, let us note that all extremal strategies of these full-correlation functions are deterministic function of the joint inputs $\vecx$, i.e., they are defined by specifying for each given $\vecx$, the corresponding sum of outputs modulo $\ell$. In other words, for each of these extremal strategies and for each given $\vecx$, we have that
\begin{equation}\label{Eq:ExtremalFullCorrelator}
	P([{\bf a}_\vecx]_k = r)=\delta_{r, f(\vecx)},
\end{equation}
where $f(\vecx)$ is some deterministic, $r$-value function of $\vecx$. Different extremal strategies of $\S_n$ in this space then corresponds to different choices of $f(\vecx)$. To prove Theorem~\ref{Thm:Simulation}, it is then sufficient to find a non-signaling strategy that gives Eq.~\eqref{Eq:ExtremalFullCorrelator} for an arbitrary choice of $f(\vecx)$.

Let us first illustrate how this works in the scenario of $n=2$. Consider the following normalized probability distribution
\begin{equation}\label{Eq:NSStrategy}
	P(a_1a_2|x_1x_2)=\frac{1}{\ell}\delta_{a_1+a_2\,{\rm mod}\, \ell\,,\, f(x_1,x_2)}.
\end{equation}
Note that (regardless of $x_1$ and $x_2$) for each $a_1$ --- due to the Kronecker delta --- there is  one, and only one value of $a_2$ such that the right-hand-side of Eq.~\eqref{Eq:NSStrategy} is non-vanishing; likewise for $a_2$. As a result, the corresponding marginal distributions are given by:
\begin{equation}\label{Eq:NSMarginals}
\begin{split}
	P(a_1|x_1x_2)=\sum_{a_2} \frac{1}{\ell}\delta_{a_1+a_2\,{\rm mod}\, \ell\,,\, f(x_1,x_2)}= \frac{1}{\ell},\\
	P(a_2|x_1x_2)=\sum_{a_1} \frac{1}{\ell}\delta_{a_1+a_2\,{\rm mod}\, \ell\,,\, f(x_1,x_2)}= \frac{1}{\ell}.
\end{split}	
\end{equation}
Both  these marginal distributions are independent of the input of the other party and hence the distribution given in Eq.~\eqref{Eq:NSStrategy} satisfies the non-signaling constraints. From these observations and Eq.~\eqref{Eq:Dfn:FullCorrelators}, it is also easy to see that the non-signaling distribution given in Eq.~\eqref{Eq:NSStrategy} satisfies Eq.~\eqref{Eq:ExtremalFullCorrelator}.
We have thus shown that in the above-mentioned subspace of full-correlation functions, the extremal strategy of $\S_2$ can also be achieved by a non-signaling correlation. More generally, for arbitrary $n\ge 2$, it is easy to verify that the following distribution:
\begin{equation}\label{Eq:NSStrategyGeneral}
	P(\veca|\vecx)=\frac{1}{\ell^{n-1}}\delta_{\sum_i a_i\,{\rm mod}\, \ell\,,\, f(\vecx)}
\end{equation}
is non-signaling, giving a uniform $n''$-partite marginal distribution of $\ell^{-n''}$, and satisfies Eq.~\eqref{Eq:ExtremalFullCorrelator}. In other words, we have proved  that the extremal strategy of $\S_n$ in the subspace of full-correlation functions can always be achieved using a non-signaling strategy.
\end{proof}

\section{Proof of Corollary~\ref{Cor:SameBound}}
\label{App:ProofCorollary}

Here, we give a proof of Corollary~\ref{Cor:SameBound}. For concreteness, we shall provide a proof for $\R=\S$. The case for $\R=\T$ follows from the inclusion relations given in Eq.~\eqref{Eq:Inclusions}.
\begin{proof}
Given inequality~\eqref{Ineq:Rn-k}, 
 the inclusion relations of Eq.~\eqref{Eq:Inclusions} immediately imply that inequality~\eqref{Ineq:Rn-k:NS}  holds true for all $\P(\veca|\vecx)\in\NS_{n,k}$.
It thus remains to show that there also exists $P(\veca|\vecx)=P^\NS_0(\veca|\vecx)\in\NS_{n,k}$ such that the inequality~\eqref{Ineq:Rn-k:NS}  is saturated, i.e., 
\begin{equation}\label{Eq:Saturating:NS}
	\sum_{\vec{x}}\sum_{r=0}^{\ell-1} \beta^{r}_{\vecx} P^\NS_0([{\bf a}_\vecx]_\ell = r)= B^\S_{n,k}.
\end{equation}

By assumption, there exists extremal $P(\veca|\vecx)=P^\S(\veca|\vecx)\in\S_{n,k}$ such that inequality~\eqref{Ineq:Rn-k}   is saturated, i.e., 
\begin{equation}\label{Eq:Saturating}
	\sum_{\vec{x}}\sum_{r=0}^{\ell-1} \beta^{r}_{\vecx} P^\S([{\bf a}_\vecx]_\ell = r)= B^\S_{n,k}.
\end{equation}
From the definition of the full-correlation function, Eq.~\eqref{Eq:Dfn:FullCorrelators},  and the assumed $k$-producibility of the correlation, we have
\begin{subequations}\label{Eq:Steps}
\begin{equation}
\begin{split}
	P^\S([{\bf a}_\vecx]_\ell = r)&=\sum_{\veca} P^\S(\veca|\vecx)\,\delta_{[{\bf a}_\vecx]_\ell,r}\\
	&=\sum_{\veca}\prod_{i=1}^G P^\S(\veca^{[i]}|\vecx^{[i]})\,\delta_{[{\bf a}_\vecx]_\ell,r},
\end{split}
\end{equation}	
where $P^\S(\veca^{[i]}|\vecx^{[i]})$ refers to the $i$-th constituent distribution, which is at most $k$-partite. Denote the sum of the outputs  in the $j$-th group by ${\bf a}_{\vecx^{[j]}}$, we can then further rewrite $P^\S([{\bf a}_\vecx]_\ell = r)$ as:
\begin{align}
	&\prod_{i=1}^G \sum_{\veca^{[i]}}P^\S(\veca^{[i]}|\vecx^{[i]})\,\delta_{\left[\sum_j \left[{\bf a}_{\vecx^{[j]}}\right]_\ell\right]_\ell,r},\\
	=&\prod_{i=1}^G \sum_{\veca^{[i]}}\sum_{r^{[i]}=0}^{\ell-1}P^\S(\veca^{[i]}|\vecx^{[i]})\delta_{\left[{\bf a}_{\vecx^{[i]}}\right]_\ell,r^{[i]}}\,\delta_{\left[\sum_j \left[{\bf a}_{\vecx^{[j]}}\right]_\ell\right]_\ell,r}.\nonumber
\end{align}
Note that for each $\vecx^{[i]}$, due to the Kronecker delta $\delta_{\left[{\bf a}_{\vecx^{[i]}}\right]_\ell,r^{[i]}}$, there is only one term in the sum over $r^{[i]}$ that contributes non-trivially. Swapping the order of the sums gives:
\begin{align}
	&\prod_{i=1}^G \sum_{r^{[i]}=0}^{\ell-1}\sum_{\veca^{[i]}}P^\S(\veca^{[i]}|\vecx^{[i]})\delta_{\left[{\bf a}_{\vecx^{[i]}}\right]_\ell,r^{[i]}}\,\delta_{\left[\sum_j \left[{\bf a}_{\vecx^{[j]}}\right]_\ell\right]_\ell,r},\nonumber\\
	=&\prod_{i=1}^G \sum_{r^{[i]}=0}^{\ell-1}\sum_{\veca^{[i]}}P^\S(\veca^{[i]}|\vecx^{[i]})\delta_{\left[{\bf a}_{\vecx^{[i]}}\right]_\ell,r^{[i]}}\,\delta_{\left[\sum_j {r^{[j]}}\right]_\ell,r},\nonumber\\
	=&\prod_{i=1}^G \sum_{r^{[i]}=0}^{\ell-1}P^\S(\left[{\bf a}_{\vecx^{[i]}}\right]_\ell=r^{[i]})\,\delta_{\left[\sum_j {r^{[j]}}\right]_\ell,r},
\end{align}
\end{subequations}
which means that $P^\S([{\bf a}_\vecx]_\ell = r)$ factorizes into a (linear combination of) product of {\em full-correlation functions} for each group $P^\S(\left[{\bf a}_{\vecx^{[i]}}\right]_\ell=r^{[i]})$. By Theorem~\ref{Thm:Simulation},  there is no loss of generality in replacing the constituent distribution from the $i$-th group $P^\S(\veca^{[i]}|\vecx^{[i]})$ by some non-signaling distributions $P^\NS_0(\veca^{[i]}|\vecx^{[i]})$ such that they agree at the level of the full-correlation functions, i.e., 
\begin{equation}
	P^\S(\left[{\bf a}_{\vecx^{[i]}}\right]_\ell=r^{[i]})=P^\NS_0(\left[{\bf a}_{\vecx^{[i]}}\right]_\ell=r^{[i]})\quad\forall\,\,i,r^{[i]}
\end{equation}
Substituting this back into Eq.~\eqref{Eq:Steps} and then Eq.~\eqref{Eq:Saturating}, we thus obtain Eq.~\eqref{Eq:Saturating:NS} by identifying
\begin{equation}
	P^\NS_0(\left[{\bf a}_{\vecx^{[i]}}\right]_\ell=r^{[i]})
	=\sum_{\veca}\prod_{i=1}^G P^\NS(\veca^{[i]}|\vecx^{[i]})\,\delta_{[{\bf a}_\vecx]_\ell,r}.
\end{equation}
\end{proof}
An immediate consequence of the above Corollary is that any full-correlation Bell-like inequality for $\S_{n,k}$, such as those derived in Refs.~\cite{Bancal:JPA:2012,Svetlichny,GMNL}, is also valid and tight for $\NS_{n,k}$.

\section{Proof of Theorem~\ref{Thm}}
\label{App:Proof}

We now provide a proof of Theorem~\ref{Thm}. 
\begin{proof}
By assumption, the following expression holds true
\begin{equation}\label{Eq:GeneralBI2}
	I_n = \sum_{\vec{a},\vec{x}} \beta^{\veca}_{\vecx} P(\vec{a}|\vec{x}){\le} 0
\end{equation}
for all $P(\veca|\vecx)\in \R_{n,k}$, and our goal is to show that
\begin{equation}
	I_{n+h} = \sum_{\veca,\vecx} \beta^{\veca}_{\vecx} P(\veca,\vec{o}|\vecx,\vec{s})\stackrel{\R_{n+h,k}}{\le} 0,
\end{equation}
for arbitrary $h\ge 1$ and  {\em all fixed} choices of $\vec{o}$ and $\vec{s}$. We will show that this is the case by {\em  reductio ad impossibilem}.

Suppose the converse, namely, that there exists some choice of $\vec{o}$, $\vec{s}$ and $h$ such that for some $P(\veca,\vec{o}|\vecx,\vec{s})\in\R_{n+h,k}$, 
\begin{equation}
	\sum_{\veca,\vecx} \beta^{\veca}_{\vecx} P(\veca,\vec{o}|\vecx,\vec{s})> 0.
\end{equation}
By linearity of the expression and the requirement that $P(\veca,\vec{o}|\vecx,\vec{s})\in\R_{n+h,k}$, the above inequality implies that there exists some correlation 
\begin{equation}\label{Eq:k-prod}
	P(\veca,\vec{o}|\vecx,\vec{s})=\prod_{i=1}^G P^\R(\veca^{[i]},\vec{o}^{[i]}|\vecx^{[i]},\vec{s}^{[i]})
\end{equation}
such that
\begin{equation}\label{Eq:Implication}
	\sum_{\veca,\vecx} \beta^{\veca}_{\vecx} \prod_{i=1}^G P^\R(\veca^{[i]},\vec{o}^{[i]}|\vecx^{[i]},\vec{s}^{[i]})> 0,
\end{equation}
where $P^\R(\veca^{[i]},\vec{o}^{[i]}|\vecx^{[i]},\vec{s}^{[i]})$ refers to the $i$-th constituent distribution (from the $i$-th group), 
and as with $P(\veca,\vec{o}|\vecx,\vec{s})$, we have used $\vec{o}^{[i]}$ and $\vec{s}^{[i]}$ to indicate, respectively, the (possibly empty) outcome and setting string that are {\em fixed} in $P^\R(\veca^{[i]},\vec{o}^{[i]}|\vecx^{[i]},\vec{s}^{[i]})$. Note  that the assumption of $P(\veca,\vec{o}|\vecx,\vec{s})\in\R_{n+h,k}$ implies that each constituent distribution is at most $k$-partite and their respective size $n_i$ sum up to $n+h$, i.e., $\sum_{i=1}^G n_i=n+h$. 

Evidently, since inequality~\eqref{Eq:Implication} is {\em strict} and that $P^\R(\veca^{[i]},\vec{o}^{[i]}|\vecx^{[i]},\vec{s}^{[i]})\ge0$ for all $\veca^{[i]}$ and $\vecx^{[i]}$, it must be the case that 
\begin{equation}\label{Eq:Marginal}
	\sum_{\veca^{[i]}}P^\R(\veca^{[i]},\vec{o}^{[i]}|\vecx^{[i]},\vec{s}^{[i]})>0
\end{equation}
for all $\vecx^{[i]}$ that contribute nontrivially in the left-hand-side of Eq.~\eqref{Eq:Implication}. In fact, since the left-hand-side of inequality~\eqref{Eq:Marginal} can also be obtained by performing the appropriate sums of Eq.~\eqref{Eq:k-prod}
\begin{align}
	\sum_{\veca,\vec{o}^{[j]} |j\neq i} P(\veca,\vec{o}|\vecx,\vec{s})&=\sum_{\veca,\vec{o}^{[j]} |j\neq i}\prod_{\ell=1}^G P^\R(\veca^{[\ell]},\vec{o}^{[\ell]}|\vecx^{[\ell]},\vec{s}^{[\ell]})\nonumber\\
	&=\sum_{\veca^{[i]}}P^\R(\veca^{[i]},\vec{o}^{[i]}|\vecx^{[i]},\vec{s}^{[i]}),
	\label{Eq:Marginal3}
\end{align}
we see that by the non-signaling nature of $P(\veca,\vec{o}|\vecx,\vec{s})$, the very last expression of Eq.~\eqref{Eq:Marginal3} must also be independent of $\vecx^{[i]}$. Hereafter, we shall simply write these marginal distributions as:
\begin{equation}\label{Eq:Marginal2}
	P^\R(\vec{o}^{[i]}|\vec{s}^{[i]})=\sum_{\veca^{[i]}}P^\R(\veca^{[i]},\vec{o}^{[i]}|\vecx^{[i]},\vec{s}^{[i]}).
\end{equation}

Hence, from inequality~\eqref{Eq:Marginal}, we see that the conditional distributions 
\begin{equation}\label{Eq:CondDist}
	\tilde{P}^{|\vec{o}^{[i]},\vec{s}^{[i]}}(\veca^{[i]}|\vecx^{[i]})=\frac{P^\R(\veca^{[i]},\vec{o}^{[i]}|\vecx^{[i]},\vec{s}^{[i]})}{P^\R(\vec{o}^{[i]}|\vec{s}^{[i]})}
\end{equation}
are well-defined for all $\vecx^{[i]}$ and satisfy the normalization condition $\sum_{\veca^{[i]}}\tilde{P}^{|\vec{o}^{[i]},\vec{s}^{[i]}}(\veca^{[i]}|\vecx^{[i]})=1$. With some thought, one can also see that the conditional distribution defined in Eq.~\eqref{Eq:CondDist} also inherits the property of the defining distribution, i.e., satisfying the constraint defined by $\R$. For instance, if $P^\R(\veca^{[i]},\vec{o}^{[i]}|\vecx^{[i]},\vec{s}^{[i]})$ admits a quantum representation, so does $\tilde{P}^{|\vec{o}^{[i]},\vec{s}^{[i]}}(\veca^{[i]}|\vecx^{[i]})$.

Dividing inequality~\eqref{Eq:Implication} by $\prod_i P^\R(\vec{o}^{[i]}|\vec{s}^{[i]})$ and using Eq.~\eqref{Eq:CondDist}, we obtain
\begin{equation}\label{Eq:Implication2}
	\sum_{\veca,\vecx} \beta^{\veca}_{\vecx} \prod_{i=1}^G \tilde{P}^{|\vec{o}^{[i]},\vec{s}^{[i]}}(\veca^{[i]}|\vecx^{[i]})> 0.
\end{equation}

As mentioned above, for all $i$, the conditional distribution $\tilde{P}^{|\vec{o}^{[i]},\vec{s}^{[i]}}(\veca^{[i]}|\vecx^{[i]})$ is a legitimate distribution with respect to the resource $\R$ and cannot be more than $k$-partite, i.e, $ \prod_{i=1}^G \tilde{P}^{|\vec{o}^{[i]},\vec{s}^{[i]}}(\veca^{[i]}|\vecx^{[i]})\in\R_{n,k}$. Hence, inequality~\eqref{Eq:Implication2} implies that the original inequality $I_n$ can be violated by correlation in $\R_{n,k}$, which contradicts our very first assumption that $I_n$ is a legitimate Bell-like inequality for $\R_{n,k}$.

\end{proof}

\end{document}